\documentclass[aps,psfig,pra]{revtex4}

\usepackage{epsfig}

\def\duzomniejsze{<\kern-.7mm<}
\def\duzowieksze{>\kern-.7mm>}

\def\textbf#1{{\bf #1}}
\def\beq{\begin{equation}}
\def\eeq{\end{equation}}
\def\be{\begin{equation}}
\def\ee{\end{equation}}
\def\ben{\begin{eqnarray}}
\def\een{\end{eqnarray}}
\def\beqa{\begin{eqnarray}}
\def\eeqa{\end{eqnarray}}
\def\eea{\end{array}}
\def\bea{\begin{array}}
\newcommand{\bei}{\begin{itemize}}
\newcommand{\eei}{\end{itemize}}
\newcommand{\bee}{\begin{enumerate}}
\newcommand{\eee}{\end{enumerate}}

\def\>{\rangle}
\def\<{\langle}

\begin{document}

\title{Quantum information isomorphism: beyond the dillemma of  
Scylla of ontology and Charybdis of instrumentalism}

\author{Ryszard Horodecki$^{(1)}$, Micha\l{} Horodecki$^{(1)}$
and Pawe\l{} Horodecki$^{(2)}$}

\affiliation{$^{(1)}$Institute of Theoretical Physics and Astrophysics,
University of Gda\'nsk, Poland}
\affiliation{$^{(2)}$Faculty of Applied Physics and Mathematics,
Gda\'nsk University of Technology, 80--952 Gda\'nsk, Poland}

\begin{abstract}
In order to have the most safe way of dealing with unanalysable quantum whole the 
Copenhagen interpretation takes as a "frame of reference" the preparation parameters
and outcomes of the measurements. It represents {\it passive} Ptolemean-like instrumentalism 
directly related to "what we see in the sky" i.e. to the "surface" of the things. 
However the notion of quantum information leads to {\it active} Copernican-like 
realism which involves (intrinsic) ordering principle and thinking about the whole 
as being analysable. One dares then to consider subsystems as localised in space, 
controlled individually, and communicated. This makes natural treating quantum information 
(quantum states) as by no means merely knowledge. Moreover it involves complementarity 
between local and nonlocal information. To avoid dilemma between Scylla 
of ontology and Charybdis of instrumentalism, the concept of 
{\it quantum information isomorphism} is proposed according to which quantum description 
of Nature and their mathematical representation. By definition it is not only just 
one-to-one mapping, but it preserves the structure. It allows, in particular, 
to treat the "wave function" as isomorphic image of what we are processing in laboratories 
implying that quantum information is indeed carried by the quantum systems. 
\end{abstract}

\maketitle

\section{Introduction}

 %
In science no one is lonely island. There are many entangled paths leading
to truth about nature. On the one of them we met Charles Bennett - 
co-discoverer of quantum information phenomena which had a decisive 
influence on development of quantum information theory
\cite{BennettBrassard,BennettWiesner,teleportation,BBPSSW1996,BBPS1996,BDSW1996}. 
During the visit at IBM we got to know his renaissance and stimulating personality. 
It is a great pleasure to one of us participate in the IBM 
Symposium honouring 60th birthday of Charles.

The aim of this contribution is to point out that the quantum information revolution 
had and still has also great influence on the change of our 
thinking about quantum formalism and its relation to physical reality.
The quantum information theory (QIT) is definitely new approach, which has significant
advantage - it allows to ask quit new questions that wouldn't be though in
the old paradigm. This new way of thinking is not only more fruitful from pragmatic
point of view. We hope it would also lead to new {\it physical} picture of nature, 
in like manner Copernican scheme was not only convenient change of frame reference
but had led us to gravitation low governing planetary motion.  

We do not want to discuss the problem 
of measurement in quantum mechanics. So far the word "interpretation
of quantum mechanics" was inevitably connected with this problem. 
After decades, the discussions become less and less fruitful, and
more and more tiresome. The main reason is that they had no 
connection with experimental results. Thus there was no criterion 
for validity of any interpretations. The Copenhagen one in a sense
won the competition, just because it was minimal. There is a 
question: is it of use to undertake any new intepretational effort? 
We believe it is still important task. Yet the goal would not be to 
solve measurement problem, but to provide a fresh view of quantum
mechanics based on the new questions that are asked within QIT.
And vice versa, one has to promote such interpretation that would 
stimulate the search a deeper ordering principle of quantum mechanics. 
QIT provided a powerful notion of quantum information, which could be 
taken as a new orientation sign, in the permanent interpretative chaos, 
rather than the minimality proposed by Copenhagen principle.

\section{Ptolemy-like instrumentalism and Copernicus-like realism
in the description of quantum phenomena}

There are historical reasons for which an instrumental Copenhagen interpretation
does not involve explicitly entanglement being at the root of the quantum
formalism. But we know it is physical property of the compound
system, then can not be ignored in building of any consistent interpretation.

The other even deeper reason for which the Copenhagen interpretation
becomes in efficient is that there was no notion of quantum information. 
The information was treated only in classical terms as knowledge
rather than the property of physical system. In fact the heart of this 
interpretation is passive Ptolemy paradigm, which takes as a reference frame the 
preparation parameters and outcomes of the measurements, that is the
"surface" of the things.

This is similar to the Ptolemy description that is related directly to what we 
see in the sky. However while looking at the sky it is hard to notice the order 
of planetary orbits as discovered by Copernicus. Yet Ptolemy description was 
artificial.  


This was known in time of Copernicus. Actually, while studying in 
Cracovian Academy, where he learned Ptolemean  
description. While carrying out the course on astronomy the 
lecturers criticised the Ptolemean approach. 

The main point 
was that the description based on just a geometrical trick made 
to obtain the picture agreeing with observations. It is important 
that both descriptions where compatible with observations. Yet Ptolemean 
approach was passive. Copernicus found that the ordering principle 
is somehow connected with Sun. This allowed Kepler and then Newton to find the deeper
 principle governing planetary motions.  We can conclude that 
Copernican approach was more isomorphic to Nature than Ptolemean. 
Therefore the former one carried more contents than it was supposed to do.

In quantum mechanics we do have Schrodinger equation, that describes 
"intrinsically" the dynamics of quantum. In Copenhagen approach, 
however this description is thought was used merely to predict the
 results of outcomes given preparation. Thus we were in situation,
 as we would be using Copernicus description just to the goal of 
describing planetary motions, instead of making step forward as 
Newton did. One of the reason is that the opposite view (de Broglie-Bohm
 and  Einstein  approach) was too classical, hence not suitable 
to reflect the curious features of quantum mechanics. Thus the 
Copenhagen interpretation was the most safe way of dealing with 
 the "great smoky dragon"  - the thing that is between preparation 
and measurement.   Quantum information theory shows that one can avoid
 two extremes: its results are not dependent on interpretation, yet the 
way of thinking is Copernican, or post-Copernican. In most cases, one
 does not think about the outcomes of measurement at all (this 
corresponds to analysing Copernican picture rather than looking at the sky).  
This gives the hope that some new organisation principle will be 
found, basing on the notions like quantum systems, the states of 
the systems, functions of the states such as entanglement, quantum
 processing the states. Such notions that can't be used in Copenhagen approach. 

In Copenhagen interpretation, we describe the experiment, not the 
quantum systems.

\begin{figure}[!h]

\psfig{file=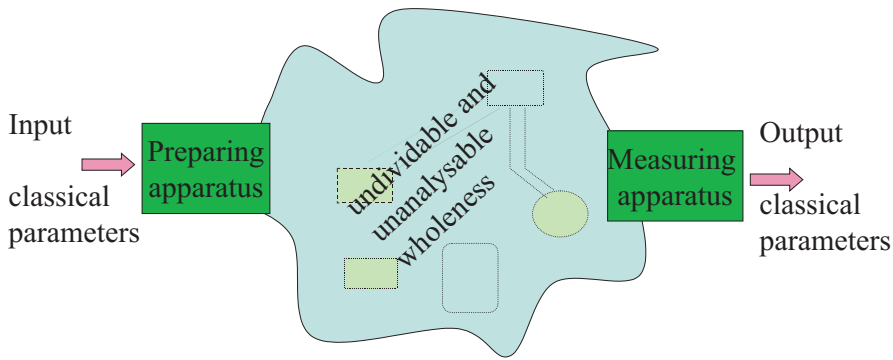, width=12cm}
\vskip-3cm
\caption[lipa]{Ptolemy-like passive paradigm takes "reference frame" the preparation 
parameters and outcomes of the measurements i.e. the "surface" of the things. 
One shouldn't think that one deals with quantum systems that are processed.
Rather apparatus includes every thing and this is unanalysable whole.}
\label{fig1}
\end{figure} 

The experiment consist of the apparatus (heavy basis)
 which is tuneable  (classical input) and which produces outcomes
 (classical output). One shouldn't think that one deals with quantum 
systems, that are processed. Rather the apparatus includes everything, 
and this is unanalysable whole. In quantum information, we think about 
the whole as analisable.
\begin{figure}[!h]
\psfig{file=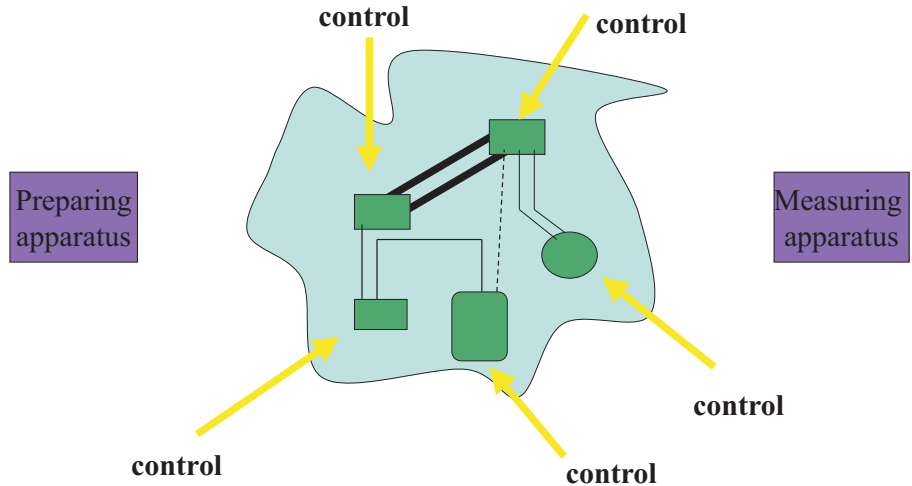, width=12cm}
\caption[lip]{Copernican-life active paradigm takes "reference frame" what is 
actually processed in laboratory. It is considered as a compound system
the subsystems of which, can be localised in space controlled individually
and communicated.}
\label{fig2}
\end{figure}

 Three stages are distinguished: 

\bei
\item preparation
\item control
\item measurement
\eei

The main difference is made by here by introducing the control stage as 
autonomous part. In that stage one has a quantum system, compound of subsystems.  
The subsystems can be localised in space. They can be controlled individually,
 and communicated. Especially the communication of quantum systems (or states) 
is completely "orthogonal" to Copenhagen approach. This quantum communication 
allows thinking about quantum information, which is by no means knowledge! We 
will elaborate it later in the context of cryptography and quantum computation.
 For the sake of present discussion, let us note that in this new
 picture offered by QIT approach, we have got the conceptual tools
 that will allow us to push forward new way of understanding Nature,
 by asking new questions.  One of such questions  is for example 
capacity of quantum channel to transmit quantum information. A simpler
 question would be: {\it can one transform one state into another one
 by means of given class of control operations? } This is the simplest
 and one of the most fundamental questions of QIT. The translation of
 this question into preparation-measurement language We leave to the 
reader as the exercise. 

Such questions can be formulated in terms of preparation-measurement 
language, yet they become completely artificial. Similarly, Newton 
wouldn't ask the question what forces the planets to go around the 
Sun, which is not natural in Ptolemean picture. In this way any
 question motivated by QIT can be rephrased in measurement-preparation
 terminology, yet it would never arise within the latter picture. In
 this way, QIT constitutes a revolution in thinking about quantum mechanics. 
Of course, even before QIT era people thought in non-Copenhagen terms, 
yet the goals of research in quantum mechanics was somewhat dominated by 
instrumental mentality enforced by minimal Copenhagen approach. One has
 to stress here, that even in QIT we cannot fully abandon
 preparation-measurement terminology. It is because it binds 
the Platonic world of wave function with what we observe in labs.
 More precisely, the preparation part can be almost completely 
absorbed into control part: e.g. in quantum computation one has 
just to prepare standard input $|00...0\>$. Just the experiment 
can be thought (on the conceptual level) as being mostly 
"control-measurement" setup. For example the quantum computer,
 if is to be useful for people, has to produce some desired 
classical output. Thus the main effort of constructors of 
quantum algorithms is the desirable connection between processing
 quantum states (Copernican part) and the  "surface" 
(what we see in the sky). Yet, again, one shouldn't be concentrated
 just on getting desirable classical input. Indeed, an important 
branch of "quantum computer science" is devoted to quantum
 input - quantum output algorithms. They are not to solve given 
classical tasks, but they solve some quantum analogues. This 
subdomain of QIT is perhaps the most contradicting the spirit
 of Copenhagen approach.

\section{Informational isomorphism}

There is a lot of discussions on what is the status of wave function.
 According to the Copenhagen approach it is a not an immanent state of
 quantum system but it describes our mathematical representation of 
knowledge about the experimental setup.  In realistic interpretations 
(in its extreme version) it is "real wave physically present in space".
 The latter approach tends to more naive realism. On the other hand the
 former approach seems to be too passive. It seems that the suitable
 approach that would not go into those extremes. Namely, any description
 of Nature can be thought as a sort of isomorphism between the 
laws Nature and their mathematical representation. By definition, 
isomorphism should be not only just one-to-one function, but it
 should preserve structure. If we insist that the role of wave
 function is merely to describe probabilities, we resign to 
treat the wave function as isomorphic image of what we are
 processing in laboratories. Using "isomorphism" approach,
 we can further claim, that quantum information is indeed 
carried by the quantum systems. The wave function is the 
image of this information. The latter cannot be written
 on paper, but the wave function can be. Thus, we would say 
that quantum information does exist, yet it is not just wave function,
 but it is represented by it. There are two main examples that support 
this view against narrow Copenhagen treatment of wave function or quantum state. 

Before discussing them let us stress that we are far away from naive realism. 
It is obvious that we can never say, for example that wave function exists. 
Yet we can {\bf argue (say)} that wave function is not merely on the paper, but rather it
is an image good mapping of something that exists. Of course we will have
never proofs that the quantum information exists. 

\subsection{ Quantum fast algorithms. }
As one knows Shor's algorithm \cite{Shor-algorytm} most probably is exponentially faster than 
any classical algorithm. If this is the case, then the wave function in 
the quantum computer at some stage of computation cannot be efficiently
 computed by classical devises. Thus people will never write this wave 
function on paper. If we insist to interpret quantum state as knowledge, 
we have to say that from operational point of view the quantum computer
 is not in any state, as nobody will know it in a reasonable time. Yet,
 something is happening during performing the algorithm, as after
 completing it, the classical outcomes give solution of the required task.
 However what we process is definitely not anybody knowledge. Now, instead
 of saying quantum computer is "great smoky dragon" we prefer to say that 
we process just the quantum information that is objectively carried by the
 system. The wave function is an image of this information, and sometimes
 we will not have wave function but the information is there. It is 
tempting to say here that we process statistics. However the statistics
 has to be encoded into quantum systems, and only this form of
 processing give powerful results such as fast factoring. Thus
 another name of quantum information is a form of encoding statistics. 
There is a notion introduced in \cite{RHkonf2,RHAnn} that is especially suitable
 to describe the paradigm we advocate here. It is "information determinism".
The quantum information carried by quantum states is processed 
deterministically because the state into which the statistics is 
encoded evolves deterministically. 
The final stage of computation is 
then to decode the statistic from the quantum information by the 
measurement. To summarise, exponentially fast quantum algorithms 
rise doubts whether treating wave function as representation 
of knowledge about the setup is fully justified. 
Or, more precisely, in this case it is inadequate 
to say that quantum evolution (algorithm) 
is evolution of our knowledge, whatever it means.  
Rather it is representation of quantum information - 
isomorphic image of the fundamental objective property 
of the system - that evolves during quantum computing.  

\subsection{ Quantum cryptography }
Quantum cryptographic key distribution schemes \cite{BennettBrassard}
(see also pioneering work \cite{Wiesner}) 
also suggest that there is a feature that is naturally ascribed to quantum systems, 
and that feature we can choose to call quantum information. 
It is clear that there must be quantum communication between
 Alice and Bob to achieve secret key. Indeed, Alice cannot simply 
throw a wave function written on paper to Bob, as Eve could read 
it out without disturbing it (knowledge is something classical, 
that can be copied). Yet, by sending quantum systems Alice and Bob
 can achieve something impossible within classical world. The goal 
of the game is to obtain a joint distribution of outcomes of Alice,
 Bob and Eve, having a feature that Alice and Bob are correlated with 
themselves, but not with Eve. Thus it must be expressed in terms of 
preparation-measurement. However the heart of the phenomenon can be 
explained in one sentence, in "Copernican" approach: quantum 
information cannot be cloned. Thus it is reasonable to think that 
Alice sends to Bob systems that carry quantum information. Again, 
he wave function represents this information, rather than being 
merely the tool to calculating the fact that finally the outcomes 
of Alice and Bob will be correlated with themselves but not with Eve's outcomes.
\begin{figure}[!h]
\vskip-1.5cm
\psfig{file=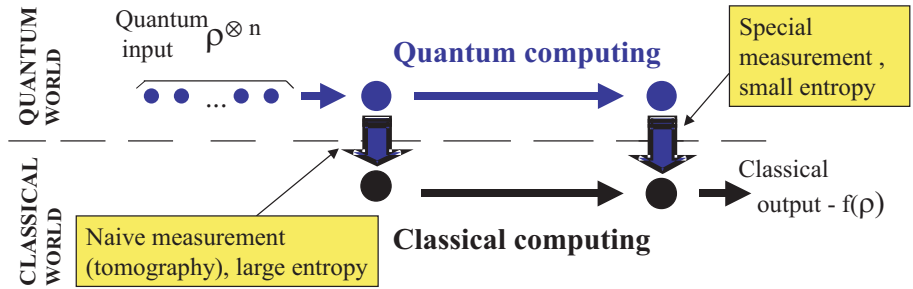, width=12cm}
\vskip-3cm
\caption{Finding function of quantum state via quantum computing. 
Quantum computing estimation of $f(\rho )$ is more natural if $\rho $ 
viewed to represent autonomic quantity to be processed rather than state of knowledge
(orthodox Copenhagen).}
\label{fig3}
\end{figure} 
\subsection{Direct computing functions of quantum state}
There is another interesting quantum scenario where both natural character of 
quantum information and its isomorphism to reality in Nature
is doubtless. This is direct calculation of quantum state 
function \cite{Estimator,PHAE}. 
Cosider a quantum stationary source that 
produces us copies of d-level system in unknown quantum state
$\varrho$. Suppose the goal is to find given function 
of quantum state $f(\varrho)$. There are two basic 
approaches. In first one performs tomography 
estimating $d^2-1$ parameters and reconstructing 
the {\it density matrix } $\varrho_{mn}$ in some basis.
Then the function of the state is  calculated 
as a function those parameters.

The second approach \cite{Estimator} is quite different - 
instead of performing estimation of many states we subject
some specially designed quantum dynamics 
that involves interaction 
between different copies of the system
and, possibly, interaction of our system with 
some controlling ancilla.
This is {\it a kind of quantum computing 
on purely quantum input}. 
Finally we subject the part of the system 
(or ancilla) to elementary binary   
measurement on quantum output which reproduces us just $f(\varrho)$.
Such approach  has two advantages:
(i) it is more natural if $\varrho$ is viewed as 
representing a sort of real quantity to be processed 
rather than state of our knowledge where the letter 
approach comes form orthodox Copenhagen interpretation.
Indeed more realistic approach to $\varrho$ 
makes it much more easy to come up with idea of such 
quantum computing schemes (not only because then
they are naturally interesting).
(ii) it is much more ''ecological''  - 
it involves much less entropy production in devices and 
records. 
We have not informational by-products in the sense 
that we do not need to collect and processed classical information 
that is to be finally (almost) entirely discarded  as {\it being  
non-interesting for observer}.

A nice example has been proposed \cite{Estimator} where the function 
\begin{equation}
f(\varrho)\equiv Tr(\varrho^k)
\end{equation}
Instead of tomography a classical matrix multiplication 
one can perform controlled shift operation 
on $k$ copies.
Final binary measurement of polarisation of controlling qubit 
reproduces the above function 
completely.

\section{Information and entanglement}
Inspired by connections between information and thermodynamics, and
on the other hand by entanglement manipulations theory we applied 
the Bennett-Landauer paradigm \cite{Landauer,Bennett82,Bennett89,LandauerPLA}
to quantum distributed systems. We considered  
work that can be drawn locally from local heat baths
by use of bipartite state that can be processed by means of 
local operations and classical communication
\cite{nlocc}, see also \cite{Zurek-demons-02}. Since work is
equivalent to information, one can think of localising information 
by LOCC. It turned out that analysing what it local (or what can be
localised) we also got to know what is non-local. Part of the 
information contents of the state is from the beginning local. 
The other part, mutual information represents correlation, and 
it can be partially localised. The part of correlation that cannot
 be localised, must be somehow quantum. For pure states we have
 obtained that the non-localisable part of information corresponds
 exactly to entanglement. Thus analysing what is local, we have 
arrived at entanglement. This binds the thermodynamical approach 
where the information losses are counted with entanglement theory. 
It is interesting, that the for two components of pure
states - localisable information and entanglement, we 
have unique measures. Indeed, we have two theories of information. 

\bei
\item NO model 
\item Pure state entanglement theory
\eei

In general the theory is given by a class of operations. Any function
 that cannot increase under the class is a resource. For reversible 
theories, there is only one type of resource. 

In NO model \cite{nlocc}, we consider simple (not compound quantum systems),
and the class of operations is noisy operations including

\bei
\item unitary operations
\item partial trace
\item adding ancilla in maximally mixed state
\eei

It turns out that under the following conditions, the theory is reversible, 
and the only resource is what we can call information. It is quantified by n-S. 

In theory of pure state entanglement \cite{BBPS1996}, the class is LOCC, and under 
the same conditions there is only one resource, quantified by entropy
of subsystem. It is called entanglement.

\section{Local-nonlocal complementarity.}

Bohr's complementarity \cite{BohrNature,BohrPhysRev} 
concerned the properties of the system 
that are observable. Indeed, Bohr thought about complementary 
classical setups of experiments. Two observables that cannot be 
measured jointly correspond to two setups of the device. In QIT
there are complementarities between quantities that are not directly
related to outcomes of measurements. Moreover, in Bohr complementarity 
there was no place for locality notion. 
In \cite{compl} we proposed a complementarity that involves the notion of
locality. It describes the mutual exclusiveness of two processes, 
performed by means of state of compound quantum systems and 
classical communication between the subsystems that are distantly 
localised. One process (i) is to gain maximal amount $I_{l}$ of local systems
in pure state. The second process (ii) is to communicate maximal amount
of quantum information $Q_{D}$. Now let us observe that instead of 
performing one of the above two tasks optimally
one can consider protocol ${\cal P}$ which produces some 
(may be nonoptimal) number  $I_{l}({\cal P})$ 
of pairs in local pure state and leads also 
to transfer of some number of qubits with the rate  
$Q_{D}({\cal P})$.
The protocol ${\cal P}$  performs two tasks
but perhaps none of them optimally.
It can be shown \cite{compl} that the following basic 
inequality holds:
\begin{equation}
Q_{D}({\cal P})+I_{l}({\cal P})\leq I_{l}
\end{equation}

%
%
%
%

One can think that information complementarity is trivial in a sense 
of the well known saying "one cannot have a cake and eat it at the
 same time". Indeed it seems natural to interpret our complementarity
 as follows: one cannot keep singlet for teleportation and at the same
 time  to use it to obtain local information. This is of course not
 complementarity, and moreover it is completely trivial. However 
our phenomenon is not trivial, for the following reasons. Firstly, 
we do not consider trade-off between singlets and local information. 
Rather we consider trade-off between the following processes: 
\bei
\item how much qubits one can teleport through the singlet
\item how many pure local qubits one can get from it a the same time
\eei

In fact it could happen that using one singlet one could teleport
 one qubit and also use the remaining state to draws 1 bit of local 
information, or at least some nonzero amount of local information. 
It would be so if in teleportation process did not produce two bits
 of entropy. However it has been shown in original paper on
 telerpotation \cite{teleportation} that one has to send two bits in teleportation
 process. This follows from causality. This fact is not obvious, 
but it is implication of quantum formalism.

So far we have shown that the phenomenon is non-trivial, yet it
 still can be called just trade-off, rather than complementarity.
 However there is another feature of our phenomenon that allows 
calling it complementarity. Namely if one decides to perform one
 task - teleportation of qubit, then one irreversibly looses the
 possibility to draw local information. And vice versa, if one 
chose to draw one bit of local information, one irreversibly 
looses possibility of teleportation of one qubit. As a matter
 of fact, one can think that after drawing local information 
one could reverse the process (investing now the information 
that has been drawn) to regain the possibility of teleportation. 
However we see that it is hopeless, because drawing local 
information destroys entanglement which irreversible process in LOCC paradigm.  

\subsection{Local-Nonlocal complementarity out of basic principles}


There is a question how fundamental is the presented complementarity.
More precisely, is our complementarity an implication of 
say, properties of space-time and bare quantum formalism, or 
does it base on some additional, less fundamental assumptions?
At first glance it seems that it bases on some specific 
additional, less fundamental assumptions. Namely it could be argued 
that more or less arbitrarily we allow only for classical communication 
between  distant observers. Such an assumption is not of fundamental nature: 
the privileged role of classical communication comes from decoherence.
The latter is on the consequence of specific form of 
physical Hamiltonians. Therefore distinction of classical communication 
cannot be regarded as a consequence basic 
properties of space-time and quantum formalism. 

However we will now argue that our complementarity is indeed a 
more fundamental phenomenon. To this end, we cannot use LOCC as  
basis.  To begin with, note that the notion of entanglement 
of pure states can be regarded as a consequence of locality and quantum 
description of compound systems. Indeed, since quantum interactions are local, 
in the sense that they decrease as the distance increase, then for 
spatially separated systems, local operations  are natural. Since in quantum 
formalism, basic role is played by pure states, it is enough to define 
entanglement for pure states. We can do so by calling the pure state 
entangled, if it cannot be produced by local operations. The concept 
can be then naturally extended to mixed states,  still without 
introducing a priori notion of classical communication, but simply by taking 
into account probabilistic nature of mixed states. A suitable 
definition of entanglement of mixed states is of course the 
one of Werner (the state is entangled iff it is not 
a mixture of product states).  

Now the notion of entanglement itself will induce a class of operations,
Namely we can single out those operations that do not create 
entanglement out of separable states (such a class was 
considered in \cite{cloning}). The operations 
are not taken a priori, but are generated by the notion of entanglement,
which in turn is a consequence of locality and quantum description of 
compound systems.  We will call the operations SP (separability 
preserving).

Let us now consider another notion - information. 
Again, as argued in \cite{uniqueinfo} it can be obtained from 
properties of basic quantum operations (unitary transformations,
adding ancilla, removing ancilla) and requirement 
of having nontrivial theory of state transformations (i.e. that 
not all states can be obtained for free). 

Having then two notions, information and entanglement, and associated 
two classes of operations, those not creating entanglement, 
and those not creating information (noisy operations NO 
\cite{nlocc}),  we  can consider a class that is intersection 
of the two: the class that does not create information and 
also does not create entanglement (call this class N(oisy)SP). 
Let us repeat here, that  this class is not an independent notion, 
assumed a priori: we have derived it from entanglement and information. 
This class is a tool, by which we can trace the flow of information 
while using the SP class. Now we can ask  how much information can 
be localised (i.e. brought to product form - again  the keyword 
locality is invoked) by SP (to this end we of course need 
to consider the joint class NSP). In this way we arrived at the notion 
of localisable information, without involving 
LOCC as basic notion.  The complementarity between 
quantum communication via SP and localisation of information 
via SP can be now viewed as a pretty fundamental one. 

Note that the class NSP is greater than the 
NLOCC one, so that the localisable information obtained here 
may be greater than the one defined by means for NLOCC. 
Thus we have modified a bit the quantity we used in derivation 
of complementarity in this paper. Yet, the modification 
does not change the main idea of interplay between 
local information 
and entanglement.

\section{Concluding remarks}
To conclude, quantum information isomorphism allowed us
to domesticate the ''great smoky dragon''- the 
reality that is between preparation and measurement.
In particular it implies informational determinism, 
according to which quantum information carried by 
quantum states is processed deterministically.
It is compatible with the generic information
paradigm, according to which the notion of information 
is fundamental category in the description 
of physical reality. On this basis existence of 
unitary information field was postulated as a necessary condition 
of any communication (or correlation).
In particular the double, hylemorphic nature of the unitary 
information field \cite{RHkonf2,RHAnn} involves two mutually coupled  
levels of physical reality: logical (informational), 
due to potential field of alternatives and energetic 
due to the fields of activities (events)
\footnote{In fact the coupling of these 
two levels was first explicitly recognised by Landauer 
\cite{Landauer} and Bennett \cite{Bennett82}.}.
The logical level naturally requires axiomatisation i.e.  
set of axioms that would extract facts that hold 
within quantum formalism and are essential 
for quantum information processing (cf. \cite{cloning}).

Finally note that quantum informational isomorphism opens 
many interesting  questions. For instance:
is there intimate connection between symmetries in Nature 
and robustness of the quantum states against quantum noise 
Is there one two one connection between fundamental interactions 
and the spectrum of physical states?
(For instance, why quarks states spectrum do not 
contain GHZ states?)  
Is gravitation somewhat distinguished in Nature? 
If so, has it influence on quantum information 
processing? Our  hope is that quantum information 
isomorphism will allow for better understanding 
of fundamental processes in nature.
Finally we would like to emphasise that quantum 
information isomorphism does not solve the 
well-known measurement problem, however 
it can serve as a guide for new interpretations 
of quantum mechanics, which would go beyond either 
of two extrema - Scylla of ontology and 
Charibdis of instrumentalism.   

Acknowledgments .- The authors would like to thank 
Christofer Fuchs, Jonathan Oppenheim and Karol 
Horodecki for discussions on foundations of quantum mechanics.

\bibliography{d:/prace/referencje/refmich}
\bibliographystyle{d:/prace/referencje/apsrev}

\end{document}